\documentclass[twocolumn, longbibliography,floatfix]{revtex4-1}

\usepackage{soul}
\usepackage{amsmath}   
\usepackage{amssymb}
\usepackage{graphicx}
\usepackage{dcolumn}
\usepackage{bm}
\usepackage{amsmath}
\usepackage{graphicx}
\usepackage{amsfonts}
\usepackage{subfigure}
\usepackage{graphicx}
\usepackage{array}
\usepackage{float}
\usepackage{color}
\usepackage{amssymb}%
\usepackage[colorlinks=true,linkcolor=blue]{hyperref}%
\hypersetup{allcolors=blue}
\usepackage[normalem]{ulem}
\usepackage{xcolor}

\newcommand{\bra}[1]{\ensuremath{\left\langle #1 \right\vert}}
\newcommand{\ket}[1]{\ensuremath{\left\vert #1 \right\rangle}}
\usepackage{mathtools}
\DeclarePairedDelimiterX\braket[2]{\langle}{\rangle}{#1 \delimsize\vert #2}

\begin{document}

\title{Non-adiabatic decay of Rydberg-atom-ion molecules}
\date{\today }

\author{A. Duspayev}
    \email{alisherd@umich.edu}
\affiliation{Department of Physics, University of Michigan, Ann Arbor, MI 48109, USA}
\author{G. Raithel}
\affiliation{Department of Physics, University of Michigan, Ann Arbor, MI 48109, USA}

\begin{abstract}
The decay of Rydberg-atom-ion molecules (RAIMs) due to non-adiabatic couplings between electronic potential energy surfaces is investigated. We employ the Born-Huang representation and perform numerical simulations using a Crank-Nicolson algorithm. The non-adiabatic lifetimes of rubidium RAIMs for the lowest ten vibrational states, $\nu$, are computed for selected Rydberg principal quantum numbers, $n$. The non-adiabatic lifetimes are  found to generally exceed the radiative Rydberg-atom lifetimes. We observe and explain a trend of the lifetimes as a function of $\nu$ and $n$, and attribute irregularities to quantum interference arising from a shallow potential well in an inner potential surface. Our results will be useful for future spectroscopic studies of RAIMs. 
\end{abstract}

\maketitle

\section{Introduction}
\label{sec:intro}
Ultralong-range Rydberg molecules (ULRM)~\cite{shafferreview, feyreview} are an active direction in Rydberg-atom research. ULRMs can be distinguished based upon their formation mechanisms. For instance, in Rydberg-ground molecules~\cite{greene,bendkowsky} a ground-state atom resides within the Rydberg-atom wave-function, and a molecular bond is formed due to scattering of the Rydberg electron at the perturber atom. The studies on this type of ULRMs include detailed analyses of association~\cite{Hamilton_2002, bendkowskyprl, tallant, bellos, andersonprl, desalvo, peper}, electronic structure~\cite{andersonpra2014}, spin-orbit coupling~\cite{kleinbach, eiles2017, deiss2020} and scattering processes~\cite{sassmann, bottcher, engel2019, jamie}, calculations and measurements of lifetimes~\cite{butscher, Butscher_2011, camargo} and permanent electric dipole moments~\cite{khuskivadze} that can vary from a few~\cite{li,bai2020} to thousands of Debyes~\cite{booth, niederprum}, and interactions with external fields~\cite{lesanovsky, hummelpra2019, hummelpra2021}. In another type of Rydberg molecules, referred to as macrodimers~\cite{boisseau,overstreet,sassmannprl}, two~\cite{boisseau} or more~\cite{samboypra2013} Rydberg atoms with non-overlapping wave-functions (LeRoy radius condition~\cite{LeRoy}) become bounded via multipolar interactions~\cite{Singer_2005, Schwettmann2006, Deiglmayr2014, marcassa}. Their formation~\cite{samboypra2011, kiffner1, Han2018, Han2019}, vibrational structure~\cite{hollerith}, lifetimes~\cite{Schwettmann2007,overstreet,sassmannprl} and alignment with external fields~\cite{hollerith2020} have been studied.

Recently, a Rydberg-atom-ion molecule (RAIM)~\cite{duspayev2021, deiss2021} has been proposed
that opens new perspectives at the interface between the fields of Rydberg molecules and atom-ion interactions~\cite{schmid, seckerpra, secker, TSchmid2018, ewald, wang, hirzler2021, dieterleprl}. In RAIMs, multipolar interaction between a Rydberg atom and an ion outside of the atom leads to bound molecular states. The non-adiabatic decay rate of RAIMs was predicted to be negligibly small~\cite{duspayev2021, deiss2021}, based upon Landau-Zener (LZ) tunneling probabilities. Since the assumptions of LZ tunneling are not satisfied in RAIMs, a quantum theory on the non-adiabatic decay of RAIMs is needed. 

Here, we develop a quantum model for non-adiabatic RAIM decay utilizing the Born-Huang representation (BHR)~\cite{BHAbook}, in which the vibrational motion is treated fully quantum-mechanically and non-adiabatic couplings are accurately described.
Being a common method to study non-adiabatic processes in conventional molecules~\cite{Agostinireview}, the 
BHR has not yet been applied to ULRMs, to our knowledge.
After reviewing the basic theory of RAIMs in Sec.~\ref{sec:raimtheory}, we describe the BHR model for their non-adiabatic dynamics in Sec.~\ref{sec:bhmodel}. Results for selected RAIMs 
are presented and discussed in Sec.~\ref{sec:results}. The paper is concluded in Sec.~\ref{sec:concl}. 

\begin{figure*}[t]
 \centering
  \includegraphics[width=0.8\textwidth]{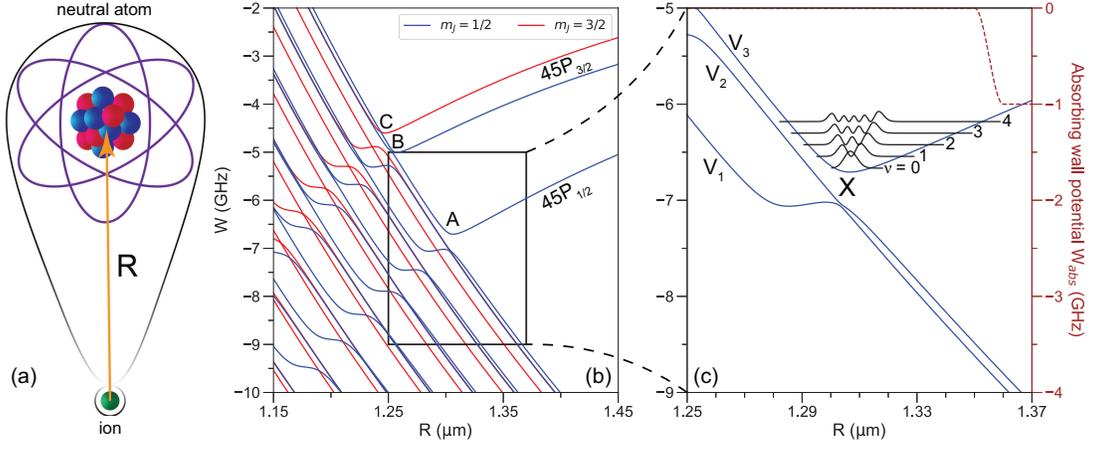}
  \caption{(a) Sketch of Rydberg-atom-ion molecule (RAIM). (b) PECs of rubidium RAIMs for $m_J=1/2$ and 3/2 as a function of internuclear distance $R$. Energies are relative to the field-free atomic $45P_{3/2}$ state. The wells in the regions A, B and C are expected to support  RAIMs. (c) Magnified view of the region A with the three $m_J=1/2$-PECs used in our calculation of non-adiabatic decay, and wave-function densities of the five lowest RAIM vibrational states in the PEC labeled $V_3$. The vertical offsets of the wave-function densities are for clarity and are not related with vibrational energy. The "X" marks the most relevant anti-crossing. The imaginary absorbing-wall potential $W_{abs}$ used in the computational approach is also shown.}
  \label{figure 1}
\end{figure*}

\section{Theory of Rydberg-atom-ion molecules} 
\label{sec:raimtheory}
The theory of RAIMs has been developed in~\cite{duspayev2021}. RAIMs, sketched in Fig.~\ref{figure 1}(a), are formed between an ion and a neutral Rydberg atom via electric-multipole interaction. The internuclear distance $R$ is larger than the radius of the Rydberg atom. Adopting a $z-$axis aligned with the internuclear axis and assuming a point-like positive ion, the interaction is, in atomic units~\cite{Schwettmann2006, Deiglmayr2014, Han2018},
\begin{eqnarray}
V_{int, m_J} (\mathbf{\hat{r}}_e; R) = - \sum_{l=1}^{l_{max}}\sqrt{\frac{4\pi}{2l+1}}\frac{\hat{r}_e^{l}}{R^{l+1}}Y_{l 0}(\hat{\theta}_e, \hat{\phi}_e).
\label{eq:Vin}
\end{eqnarray}
Here, $m_J$ is the conserved magnetic quantum number of the Rydberg atom, $n$ the principal quantum number, $\ell$ the orbital quantum number, $\mathbf{\hat{r}}_e = (\hat{r}_e, \hat{\theta}_e, \hat{\phi}_e)$ the Rydberg-electron position operator relative to the atom's center, and $Y_{l0}(\hat{\theta}_e, \hat{\phi}_e)$ a spherical harmonic in which $l$ is the multipole order of the atomic charge distribution. Diagonalization of the Hamiltonian with atom-ion interaction given in Eq.~\ref{eq:Vin} yields the molecular potential energy curves (PECs). Some PECs exhibit deep wells conducive to bound vibrational states of RAIMs, such as PECs in the vicinity of $nP_J$ Rydberg states of cesium and rubidium. The characteristics of these RAIMs depend on quantum defects and other atomic parameters~\cite{duspayev2021,deiss2021}. 

In Fig.~\ref{figure 1}(b), we show a case in which RAIMs are formed below the Rb $45P$ asymptotes. The RAIM potential wells are several hundreds of MHz deep and on the order of 100~nm wide, which leads to tens of bound vibrational states. In Fig.~\ref{figure 1}(c), we show the lowest 5 RAIM states in the molecular PEC labeled $V_3$. The stability of the vibrational states may be affected by the non-adiabatic couplings between $V_3$ and neighboring PECs, labeled $V_1$ and $V_2$ in Fig.~\ref{figure 1}(c). 

\section{Non-adiabatic dynamics} 
\label{sec:bhmodel}

\subsection{Time-dependent Schr\"{o}dinger equation}
\label{subsec:tdsebhr}

In the Born-Oppenheimer approximation (BOA)~\cite{BOA}, electronic and nuclear wave-functions are adiabatically separated to facilitate the calculation of vibrational molecular states. However, the BOA can "break down" when vibrational and electronic time scales approach each other, as is the case when PECs exhibit narrow anti-crossings, leading to non-adiabatic coupling of PECs and to molecular decay. In Fig.~\ref{figure 1}(c), non-adiabatic coupling from $V_3$ to the unbound PECs may contribute significantly to RAIM decay.

A common method to model non-adiabatic effects is to write the TDSE in BHR~\cite{BHAbook} (see~\cite{Agostinireview} and references therein for a recent overview of BHR theory and applications).
Here, we consider the vibrational degree of freedom of a diatomic RAIM along its internuclear axis,  $\hat{\mathbf{R}}$. The TDSE in BHR is written as:
\begin{eqnarray}
i \hbar \frac{\partial \psi_i (R, t)}{\partial t} = -\frac{\hbar^2}{2 \mu} \frac{\partial^2 \psi_i (R, t)}{\partial R^2} + V_i(R) \psi_i (R, t) \nonumber\\ 
+ \sum_{j} \mathcal{F}_{ij}(R) \psi_j(R,t).
\label{eq:BHSE1}
\end{eqnarray}
\noindent 
Here, $\psi_i (R, t)$ is the adiabatic vibrational RAIM wave-function on $V_i (R)$, $\mu$ is the reduced mass, and the $\mathcal{F}_{ij}(R)$ are the non-adiabatic couplings between the adiabatic wave-functions on PECs $V_i$ and on $V_j$. The non-adiabatic terms of the BHR, $\mathcal{F}_{ij}(R)$, couple nuclear and electronic motion. Explicitly,
\begin{eqnarray}
\mathcal{F}_{ij}(R) = \mathcal{A}_{ij}(R) \cdot \frac{\partial}{\partial R} + \mathcal{B}_{ij}(R),
\label{eq:ABterms}
\end{eqnarray}
\noindent where $\mathcal{A}_{ij} (R)$ is referred to as the first-order non-adiabatic coupling and is defined as (in one dimension): 

\begin{eqnarray}
\mathcal{A}_{ij}(R) = -\frac{\hbar^2}{\mu} \bra{\phi_i} \frac{\partial}{\partial R} \ket{\phi_j} \quad.
\label{eq:Aterm}
\end{eqnarray}
\noindent There, $\vert \phi_i (R) \rangle$ are the $R$-dependent electronic states of the Rydberg atom, with wave-functions
$\phi_i({\bf{r}}_e; R) = \langle {\bf{r}}_e \vert \phi_i (R) \rangle$. 
The inner product in Eq.~\ref{eq:Aterm} is evaluated in the Rydberg-electron state space, {\sl{i. e.}} it involves, in principle, an integral over 
${\bf{r}}_e$. In practice, the Rydberg state is represented in the ``diabatic'' Rydberg-state basis $\{ \vert n, \ell, J, m_J \rangle =: \vert \alpha \rangle \}$, with the shorthand $\alpha$ for all diabatic-state quantum numbers. The electronic wave-function on PEC $i$ then reads 
\[\phi_i({\bf{r}}_e; R) = \sum_{\alpha} \quad c_{i,\alpha}(R) \langle 
{\bf{r}}_e \vert \alpha \rangle \quad, \]
with coefficient functions $c_{i,\alpha}(R)$. Because of the $R$-independence and the orthonormality of the $\vert \alpha \rangle$,  
\begin{eqnarray}
\mathcal{A}_{ij}(R) = -\frac{\hbar^2}{\mu} 
\sum_{\alpha} \quad c^*_{i,\alpha}(R) \left[ \frac{\partial}{\partial R} c_{j,\alpha}(R) \right] \quad.
\label{eq:Aterm2}
\end{eqnarray}
The $\mathcal{B}_{ij} (R)$ in Eq.~\ref{eq:ABterms} is referred to as the second-order non-adiabatic coupling and is 
\begin{eqnarray}
\mathcal{B}_{ij}(R) & = & -\frac{\hbar^2}{2 \mu} \bra{\phi_i} \frac{\partial^2}{\partial R^2} \ket{\phi_j} \nonumber \\
& = & -\frac{\hbar^2}{2 \mu} \sum_{\alpha} \quad c^*_{i,\alpha}(R) \left[ \frac{\partial^2}{\partial R^2} c_{j,\alpha}(R) \right]
\quad .
\label{eq:Bterm}
\end{eqnarray}
\noindent 

Since $\mathcal{A}_{ij} (R) = - \mathcal{A}_{ji} (R)$, it is $\mathcal{A}_{ii} (R) = 0$. The remaining (generally non-zero) diagonal non-adiabatic couplings, $\mathcal{B}_{ii} (R)$, are often combined with the corresponding $V_i (R)$ into
\begin{eqnarray}
\tilde{V}_i(R) = V_i (R) + \mathcal{B}_{ii}(R) \quad.
\label{eq:Vadiab}
\end{eqnarray}
We may then re-write Eqs.~\ref{eq:BHSE1} and~\ref{eq:ABterms} as
\begin{eqnarray}
i \hbar \frac{\partial \psi_i (R, t)}{\partial t} = -\frac{\hbar^2}{2 \mu} \frac{\partial^2 \psi_i (R, t)}{\partial R^2} + \tilde{V}_i(R) \psi_i (R, t) \nonumber\\ 
+ \sum_{j\neq i} \Big[ \mathcal{A}_{ij}(R) \frac{\partial \psi_j(R,t)}{\partial R} + \mathcal{B}_{ij}(R)\psi_j(R,t) \Big],
\label{eq:BHSE2}
\end{eqnarray}
\noindent We refer to $\tilde{V}_i(R)$ as ``adiabatic potentials'', {\sl{i.e.}} potentials in which all diagonal non-adiabatic energy shifts of the adiabatic states have been added to the PECs, $V_i(R)$. We also note that for $\mathcal{B}$ it is
\begin{equation}
\mathcal{B}_{ij} + \mathcal{B}_{ji}^* = - \frac{\hbar^2}{\mu} \big[\frac{\partial }{\partial R} \langle \phi_i  \vert \big]
\big[\frac{\partial }{\partial R} \vert  \phi_j \rangle \big] 
\quad. \nonumber 
\end{equation}
We use this identity as a check for numerical errors caused  by the step size in $R$. 

It is apparent from Eqs.~\ref{eq:Aterm}-\ref{eq:Bterm} that the non-adiabatic couplings follow from the $R$-dependencies of the adiabatic Rydberg states $\phi_i({\bf{r}}_e; R)$, which are critically affected by the avoided crossings between the PECs. As the general shapes of the PECs are the same for all cases studied here, in Sec.~\ref{sec:results} we find a general trend for the non-adiabatic RAIM decay times. However, detailed differences in the non-adiabatic $\mathcal{A}$- and $\mathcal{B}$-terms and in the wave-function dynamics on the dissociative PECs lead to peculiar quantum effects that are also discussed. 

\subsection{Simulation method}
\label{subsec:simulation}

We first numerically calculate PECs for selected Rydberg states $nP_J$ of $^{87}$Rb (see, for instance, Fig.~\ref{figure 1}(b) for $45P_J$). We investigate bound RAIM vibrational states on PECs that asymptotically connect with the $nP_{1/2}$ levels.
The corresponding PEC for $n=45$ is labeled $V_3$ in Fig.~\ref{figure 1}(c). This PEC, and equivalent PECs for other $n$-values, exhibit non-adiabatic couplings mostly to a pair of lower, dissociating PECs labeled $V_1$ and $V_2$ in Fig.~\ref{figure 1}(c). The PEC calculation yields the PECs, the associated adiabatic Rydberg states $\vert \phi_i (R) \rangle$, and the non-adiabatic terms 
$\mathcal{A}_{ij} (R)$ and $\mathcal{B}_{ij} (R)$
according to the equations in
Sec.~\ref{subsec:tdsebhr}.  In view of the structure of the PEC anti-crossings evident in Fig.~\ref{figure 1}, the effect of non-adiabatic couplings of $V_3$ to PECs other than $V_1$ and $V_2$ is deemed negligible. The Rydberg-state basis sets used in the PEC calculations include all Rydberg levels with $m_J = 1/2$ and effective principal quantum number differing by less than 5 from that of the molecular RAIM states of interest. 

The computational method to solve the TDSE in Eq.~\ref{eq:BHSE2} on the three relevant PECs $V_1$, $V_2$ and $V_3$ is a Crank-Nicolson (CN) algorithm~\cite{cnaref}. The simulation is initialized with a RAIM vibrational state on the adiabatic potential $\tilde{V}_3$ (which differs slightly from the PEC $V_3$, according to Eq.~\ref{eq:Vadiab}). Over the course of the subsequent simulated evolution, the norm of the wave-function decays due to the non-adiabatic couplings, allowing us to extract the molecular lifetimes. In the following, we describe additional details of the method.      

The potential $\tilde{V_3} (R)$, constructed according to Eq.~\ref{eq:Vadiab}, is used to calculate the initial vibrational RAIM state, $\Psi_\nu (R)$, with vibrational quantum number $\nu$. The initial state for the CN simulation then is $\psi_1 (R, t = 0) = \psi_2 (R, t = 0) = 0$ and  
$\psi_3 (R, t = 0) = \Psi_\nu (R)$. The wave-function is propagated in time for a duration of $t_{total} = 50~\mu$s with a step size $\Delta t = 20$~ps. In order to reduce transients from sudden ``turn-on'' of the non-adiabatic terms, the non-adiabatic terms $\mathcal{A}_{ij} (R)$ and $\mathcal{B}_{ij} (R)$ are slowly ramped up at the beginning of the time-propagation. We still find minor initial transients in the $\psi_i(R,t)$, which cease at times $t_0 \sim 10$~ns.

The dissociating potentials, $V_1$ and $V_2$, are unbound.
As our CN simulation employs a spatial box with fixed boundary conditions $\psi_i(R) = 0$ on all boundaries,
the potentials must be modified such that wave-functions propagating outward on $V_1$ and $V_2$ are absorbed rather than reflected. To terminate the outgoing wave-function, we add an imaginary part, $Im [V_i] = W_{i, abs} (R)$, on the unbound potentials $V_1$ and $V_2$, as depicted by the dashed line in Fig.~\ref{figure 1}(c). The domain over which $W_{i, abs} (R)$ differs from zero is placed far enough out in $R$ that it does not affect the non-adiabatic dynamics of interest, which is restricted to regions within which the non-adiabatic couplings differ from zero. The absorbing wall $W_{i, abs} (R)$ exhibits a smooth turn-on, so as to avoid reflections. We have checked the effectiveness of the absorbing wall as well as the absence of wall reflections by calculating the quantum flux as a function of $R$ (outside and inside the wall), and by verifying the absence of standing-wave patters on $\psi_1$ and $\psi_2$ near the absorbing wall.

The absorbed outgoing flux leads to a decay of the overall wave-function norm~\cite{muga2004}, allowing us to extract the RAIM lifetime. The population in $\psi_3$, $p_3(t) = \int |\psi_3(R,t)|^2 dR$, is determined as a function of propagation time and fitted to the function: 
\begin{eqnarray}
p_3 (t) = p_3 (t_0) e^{-(t-t_0)/\tau_{nad}},
\label{eq:fitfunction}
\end{eqnarray}
\noindent with fitting parameters $p_3(t_0)$ and $\tau_{nad}$. Here, 
$p_3(t_0) \lesssim 1$ reflects the population after ramping up the non-adiabatic terms and after allowing transients to cease, and $\tau_{nad}$ is the non-adiabatic RAIM lifetime for the given $n$ and $\nu$. 

\begin{figure}[t]
 \centering
  \includegraphics[width=0.47\textwidth]{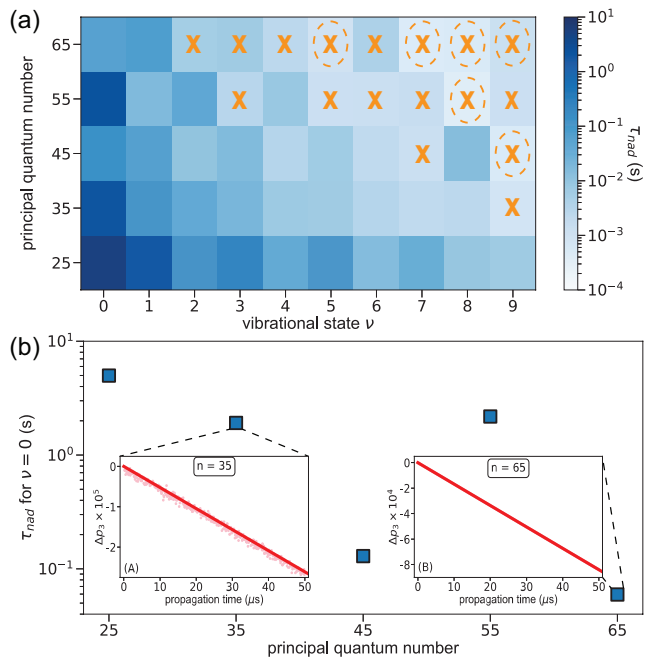}
  \caption{(a) Calculated non-adiabatic lifetimes, $\tau_{nad}$, of rubidium RAIMs below the $nP_{1/2}$ asymptotes vs vibrational and principal quantum numbers, $\nu$ and $n$, displayed on the indicated logarithmic color scale. The X-marks and dashed circles mark cases in which $\tau_{nad}$ is less than ten times the radiative decay time at 1.2~K and 300~K, respectively.
  (b) The lifetimes $\tau_{nad}$ for the ground vibrational state, $\nu = 0$, corresponding to the first column in (a). The insets show the computed 
  %\afix{differential?} 
  population decrease, $\Delta p_3 (t)$, of the ground RAIM states for $n = 35$ (inset (A)) and $n = 65$ (inset (B)) as a function of time. The $\tau_{nad}$-values follow from the slopes according to Eq.~\ref{eq:fitfunction}.}
  \label{figure 2}
\end{figure}

Although most $\tau_{nad}$-values are longer than $t_{total}$, as seen in Fig.~\ref{figure 2}, the decrease of $p_3(t)$ during the interval
$t_{total}$ allows for an accurate determination of $\tau_{nad}$ in all cases studied.
We have checked that lowering the computation time step $\Delta t$ does not significantly alter the $\tau_{nad}$.  

\section{Results and Discussion} 
\label{sec:results}

We obtain the non-adiabatic lifetimes, $\tau_{nad}$, of the lowest ten vibrational states of RAIMs below the $nP_{1/2}$ Rydberg-state asymptotes for $n \in \{25, 35, 45, 55, 65\}$. The results are listed in Table~\ref{tab:table1} and visualized in Fig.~\ref{figure 2}(a). These lifetimes are much longer than the radiative lifetimes of $nP$ Rydberg states, $\tau_{r}$, which have values between $\tau_{r}=32~\mu$s for $n=25$ and 680~$\mu$s for $n=65$ in a 1.2-K black-body radiation field. 
In 300~K radiation fields, the $\tau_{r}$-values are between 17~$\mu$s and 160~$\mu$s for $n=25$ and 65, respectively, because upward and downward bound-bound transitions as well as thermal ionization reduce the lifetime~\cite{gall}.  Our $\tau_{r}$-values,
obtained in context with work involving Rydberg-atom-state diffusion in thermal radiation fields~\cite{Traxler2013, Anderson2013},
are roughly in-line with values of $\sim 20 \mu$s at $n = 25$~\cite{Branden} and $\sim 150 \mu$s at $n = 65$~\cite{beterov2009} reported elsewhere.

Under the absence of other decay channels, the net RAIM decay time $\tau = (1/\tau_{nad} + 1/\tau_{r})^{-1}$. The symbols on the color map in Fig.~\ref{figure 2}(a) mark cases in which $\tau_{nad} < 10 \tau_{r}$ in 1.2~K and 300~K black-body fields. Figure~\ref{figure 2}(a) shows that non-adiabatic RAIM decay, while not being the dominant decay mechanism, should have a noticeable effect at higher $n$-values and vibrational quantum numbers $\nu$. We note that the values for $\tau_{r}$ assumed in Fig.~\ref{figure 2}(a) are for $nP$ Rydberg states, whereas the
electronic states of RAIMs states carry up to about 50\% admixture of longer-lived high-$\ell$ Rydberg levels. This means non-adiabatic decay might be slightly more relevant, on a relative scale, than suggested in Fig.~\ref{figure 2}(a).

\begin{table}[t]
\caption{\label{tab:table1} Calculated non-adiabatic lifetimes $\tau_{nad}$ in seconds.}
\begin{ruledtabular}
\begin{tabular}{c c c c c c}
$\nu$ & $n = 25$ & $n = 35$ & $n = 45$ & $n = 55$ & $n = 65$ \\
\hline
0 & 4.99 & 1.91 & 0.129 & 2.180  & 0.0595\\
1 & 1.67 & 0.116 & 0.0627 & 0.0175 & 0.0724\\
2 & 0.123 & 0.0393 & 0.0105 & 0.0425 & 0.00570\\
3 & 0.238 & 0.0216 & 0.0167 & 0.00297 & 0.00659\\
4 & 0.0367 & 0.00773 & 0.00324 & 0.00855 & 0.00259\\
5 & 0.103 & 0.00685 & 0.00631 & 0.00122 & 0.00103\\
6 & 0.0151 & 0.00320 & 0.00219 & 0.00135 & 0.00426\\
7 & 0.0295 & 0.00221 & 0.00132 & 0.00162 & 0.000475\\
8 & 0.00920 & 0.00262 & 0.0143 & 0.000411 & 0.000534\\
9 & 0.00702 & 0.000851 & 0.000572 & 0.00120 & 0.00139\\
\end{tabular}
\end{ruledtabular}
\end{table}

To illustrate the inadequacy of LZ estimates for non-adiabatic RAIM decay, we consider the RAIM vibrational ground state for $45P_J$, for which we have computed $\tau_{nad} \sim 0.13$~s (see Fig.~\ref{figure 2}(b) and Table~\ref{tab:table1}). From the PECs in Fig.~\ref{figure 1} and the vibrational energy levels we estimate a vibration frequency of $f_0 = 18.2$~MHz, corresponding to a LZ decay ``attempt rate'' of $R = 2 f_0=36.4$~MHz. The main avoided crossing has a gap size of $G \approx h\times$350~MHz (see gap marked ``X'' in Fig.~\ref{figure 1}(c)).  For a LZ RAIM decay estimate, we assume a fixed particle velocity given by the vibration velocity, $v_{max}$, at the minimum of $V_3$. From $(\mu/2) v_{max}^2 = h f_0/2$, with effective mass $\mu = 43.5$~amu for $^{87}$Rb, we obtain $v_{max} = 0.41$~m/s. 
The differential slope of the level crossing, estimated from Fig.~\ref{figure 1}(b), is $s = h \times 49$~GHz/$\mu$m. The LZ tunneling probability then is $P_{LZ} = \exp(-2 \pi (G/2)^2 / \hbar / s / v_{max}) = 7 \times 10^{-27}$, and the LZ RAIM lifetime $\tau = 1/ (R P_{LZ}) = 4 \times 10^{18}$~s. This estimate is about 20 orders of magnitude too large.

\begin{figure}[htb]
 \centering
  \includegraphics[width=0.5\textwidth]{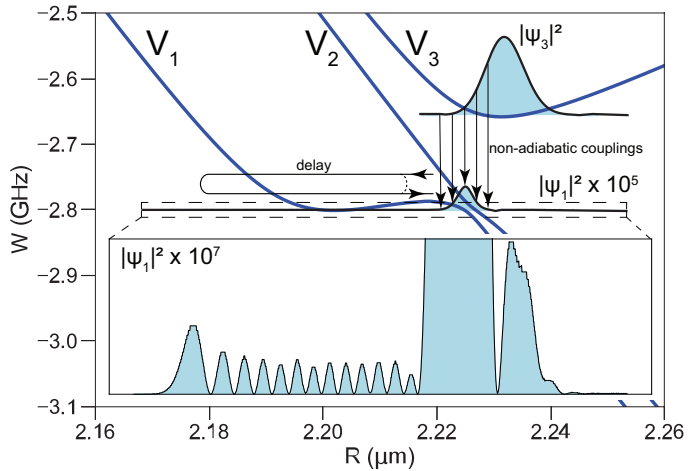}
  \caption{Wave-function densities $|\psi_1|^2$ and $|\psi_3|^2$ in the respective potentials $V_1$ and $V_3$ after reaching a quasi steady-state
  %after several time-step propagations 
  for $55P_{1/2}$. The inset shows an enlarged view of $|\psi_1|^2$.
  }
  \label{figure 3}
\end{figure}

From Fig.~\ref{figure 2} and Table~\ref{tab:table1} it is seen that the non-adiabatic lifetimes follow a downward trend with increasing $n$ and increasing $\nu$. This is expected because the anti-crossing gaps decrease with increasing $n$, and because the $\partial/ \partial R$-operator, which occurs in combination with the $\mathcal{A}$-terms, exacerbates the non-adiabatic coupling at higher $\nu$-values, where the vibrational wave-function gradients become larger. 

Inspecting Fig.~\ref{figure 2} and Table~\ref{tab:table1}, we further note substantial deviations of $\tau_{nad}$ from the overall trend. {\sl{E.g.}}, $\tau_{nad}$ for $\nu = 0$ has an outlier at $n = 55$, for which $\tau_{nad}$ is larger than even for $n = 35$. We believe that the irregularities originate in a quantum-interference effect in the RAIM decay. To illustrate this, in Fig.~\ref{figure 3} we show PECs $V_1$ and $V_3$ and their adiabatic wave-function densities. The non-adiabatic couplings set up a quasi-stationary $\psi_1$ and an associated probability-density flow that causes the non-adiabatic RAIM decay. The density $\vert \psi_1 \vert^2$ relative to $\vert \psi_3 \vert^2$ is small, but non-zero (even at the right margin of the plot), and it exhibits a standing wave in a shallow well in $V_1$ centered around $\approx 2.20~\mu$m. The standing wave is due to a 100-\% reflection on the rising side of $V_1$ near 2.18~$\mu$m, and a partial quantum reflection at the three-level crossing near 2.23~$\mu$m. 
The net outward flow on $V_1$ to the right of the anti-crossing is a superposition of a contribution due to direct non-adiabatic coupling from $V_3$ onto $V_1$, and a contribution that proceeds via non-adiabatic coupling from $V_3$ into the shallow potential well in $V_1$ and time-delayed escape, as visualized by the delay loop in Fig.~\ref{figure 3}. The superposition amplitude depends on the phase difference between the contributions, which varies as a function of $n$ and $\nu$, causing the irregularities seen in Fig.~\ref{figure 2}. A few combinations of $n$ and $\nu$ appear to exhibit substantial destructive interference, leading to lifetimes that are much longer than the overall trend would suggest.  

The situation portrayed in Fig.~\ref{figure 3} applies to all values of $n$ studied. Our interpretation of the lifetime irregularities in Fig.~\ref{figure 2} in terms of a quantum interference effect has been supported in additional test calculations, not shown, in which an absorbing potential has been placed within the shallow potential well in $V_1$. The test calculations show a smooth dependence of $\tau_{nad}$ without irregularities.  

\section{Conclusion} 
\label{sec:concl}
In summary, we have presented results of calculations of non-adiabatic decay of Rydberg-atom-ion molecules of Rb. The lifetimes, extracted for five representative values of $n$ and for the lowest ten vibrational states, follow an overall trend that is in-line with the behavior of avoided-crossing gap sizes and the structure of the vibrational wave-functions. Deviations from the trend were attributed to a quantum interference effect. Our results confirm that the RAIM states are quite stable against non-adiabatic decay, and that their lifetimes are mainly limited by radiative decay of the Rydberg valence electron. Future experimental work may reveal the existence and the lifetimes of RAIM molecules. Future computational work could be devoted to non-adiabatic decay of RAIMs in other potential wells evident in Fig.~\ref{figure 1}(b), as well as a more detailed study of quantum interference effects in RAIM decay. The formalism discussed here could also be applied to study non-adiabatic processes in other types of Rydberg molecules. 

\section*{ACKNOWLEDGMENTS}
We thank Bineet Kumar Dash and Ansh Shah for useful discussions. This work was supported by NSF grant No. PHY-1806809 and in part through computational resources and services provided by Advanced Research Computing at the University of Michigan, Ann Arbor.

\bibliography{references.bib}

\end{document}